\documentclass[letter]{aa}  

\usepackage{graphicx}
\usepackage{txfonts}

\usepackage{natbib}
\usepackage{psfrag}
\usepackage{bm}
\usepackage{amsmath}

\bibpunct{(}{)}{;}{a}{}{,} 

\def\be{\begin{equation}}
\def\ee{\end{equation}}
\def\kms{{\rm\,km\,s^{-1}}}
\def\kmskpc{{\rm\,km\,s^{-1}\,{kpc}^{-1}}}
\def\pc{{\rm\,pc}}

\def\yr{{\rm\,yr}}

\def\deg{{^\circ}}

\def\kpc{{\rm\,kpc}}

\def \sun{{_\odot}}
\def\1s{{1$\sigma$}}
\def\2s{{2$\sigma$}}
\def\3s{{3$\sigma$}}

\newcommand{\bea}	{\begin{array}}
\newcommand{\eea}	{\end{array}}
\newcommand{\ben}	{\begin{eqnarray}}
\newcommand{\een}	{\end{eqnarray}}
\newcommand{\bsq}	{\begin{mathletters}}
\newcommand{\esq}	{\end{mathletters}}

\newcommand{\vl}	{V_\ell}
\newcommand{\vb}	{V_b}

\newcommand{\Vs} 	{V_\odot}
\newcommand{\Us} 	{U_\odot}

\newcommand{\eqn}[1]	{equation\ (\ref{#1})}

\newcommand{\Sec}[1]	{Section~\ref{#1}}
\newcommand{\Fig}[1]	{Figure~\ref{#1}}
\newcommand{\fig}[1]	{Fig.~\ref{#1}}

\newcommand{\D}	{\Delta}
\newcommand{\De}	{\Delta_{\rm exp}}

\def\aap{A\&A}

\def\2s{2-$\sigma$}
\def\3s{3-$\sigma$}

\def\mas{{\rm\,mas}}

\def\gaia{\emph{Gaia }}

\begin{document}

   \title{The intricate Galaxy disk: velocity asymmetries in Gaia-TGAS }
   \titlerunning{Velocity asymmetries in TGAS}


   \author{T. Antoja
          \inst{1}
           \and
          J. de Bruijne\inst{2}
 \and
F. Figueras\inst{1}         
\and
                   {R. Mor\inst{1}  }
                   \and
                   T. Prusti\inst{2}  
\and
S. Roca-F\`abrega\inst{3}
          }

   \institute{Dept. FQA, Institut de Ciencies del Cosmos (ICCUB), Universitat de Barcelona (IEEC-UB), Marti Franques 1, E08028 Barcelona, Spain, 
              \email{tantoja@fqa.ub.edu}
         \and
             Directorate of Science, European Space Agency (ESA-ESTEC), PO Box 299, 2200 AG Noordwijk, The Netherlands
           \and
Racah Institute of Physics, The Hebrew University of Jerusalem, Edmond J. Safra Campus, Givat Ram, Kaplun building, office 110, Jerusalem 91904, Israel
             }

   \date{Received September xx, xxxx; accepted March xx, xxxx}

 
  \abstract
   {We use the Gaia-TGAS data to compare the 
transverse velocities in Galactic longitude (coming from proper motions and parallaxes) in the Milky Way disk for negative and positive longitudes as a function of distance. The transverse velocities are {strongly asymmetric} and deviate significantly from  the expectations for an axisymmetric Galaxy. The value and sign of the asymmetry 
changes at spatial scales of several tens of degrees in Galactic longitude and about $0.5\kpc$ in distance. The asymmetry is statistically significant at ${95\%}$ {confidence} level for ${57\%}$ of the region probed, which extends up to $\sim1.2\kpc$. A percentage of ${24}\%$ of the region studied shows absolute differences {at this confidence level} larger than ${5\kms}$ and ${7\%}$ larger than $10\kms$.
  The asymmetry pattern shows mild variations in
  {the vertical direction} and with stellar type. A first qualitative comparison with spiral arm models indicates that the arms {are unlikely to be} the main source of the asymmetry. We briefly discuss alternative origins. This is the first time that {\em global all-sky} asymmetries {are detected} in the Milky Way kinematics, beyond the local neighbourhood, and with a purely astrometric sample. }

   \maketitle
%

\section{Introduction}\label{intro}
The scientific community studying the Galaxy welcomed with much expectation the publication of the first \gaia data \citep{GaiaCollaboration2016b,GaiaCollaboration2016} in September of 2016. Even with the limitations of the first release, the Gaia data possesses exciting possibilities for new discoveries on the formation, evolution and current structure of the Milky Way. The Tycho-Gaia astrometric solution (TGAS, \citealt{Michalik2015}) is the largest astrometric sample to date and comprises proper motions and parallaxes of unprecedented accuracy for two-million stars.

Following the approach proposed in \citet{Antoja2016} (hereafter A16), here we compare the transverse velocities in Galactic longitude (i.e., coming from proper motions and parallaxes) for negative and positive Galactic longitudes as a function of distance with the Gaia-TGAS data. Once the solar motion is subtracted, the data reveals clear large-scale velocity asymmetries that are signatures of the non-axisymmetry in the Galaxy.

This discovery adds up to recent findings of the \textit{intricateness} of the Galactic disk. For instance, there is multiple evidence of radial and vertical velocity gradients and wave-like motion in the disk  (e.g. \citealt{Siebert2011a,Widrow2012}). Also, \citet{Carlin2013} found asymmetric vertical and radial velocities for different azimuths in the direction of the anti-centre.
Recently, \citet{Bovy2017} found evidence of non-zero values of the local divergence and radial shear while measuring the Oort's constants with TGAS, but restricted to a local sample ($\sim200\pc$).
So far these detections have been either very local, limited to the directions probed with the particular ground-based survey, or detected primarily in radial velocity data. Thanks to Gaia, this is the first time that we detect {\em global all-sky} velocity asymmetries, beyond the local neighbourhood and with a purely astrometric sample.

We describe the data in \Sec{data} and the method in \Sec{method}.  We show our results on the velocity asymmetry in \Sec{results}. We perform tests to asses the limitations and robustness of our results in \Sec{strlim}. In \Sec{conclusion} we conclude and comment on the possible origin of the detected asymmetry.


\section{The data}\label{data}

We use proper motions from Gaia-TGAS (\citealt{GaiaCollaboration2016b}, {\citealt{Lindegren2016}}) and the distance estimations from \citet{Astraatmadja2016}. These were obtained in a Bayesian way from the TGAS parallaxes using different priors: i) an isotropic prior with an exponentially decreasing space density with increasing distance with a short scale length (hereafter dist1), ii)  with a longer scale length (dist2), and iii) an anisotropic prior derived from the observability of stars in a Milky Way model (dist3). 
{We also compare the results with the inverse of the parallax as a distance estimator.} 

We select different layers in $Z$. We assume a Sun's height above the {plane} $Z\sun=0.027\kpc$ \citep{Chen2001}. Our primary sample is a disk layer with $|Z|<0.1\kpc$ and has 936861 stars. 
We further divide our primary sample into groups of different spectral types and luminosity classes obtained from  the catalogue of \citet{Pickles2010} that we cross-match with TGAS using the Tycho2 ID. We will study young main sequence stars (OBAV, 66446 stars), main sequence stars (FGKMV, 509874 stars) and giant stars (KMIII, 126988 stars). 

For these samples we compute the observed transverse velocity in the longitude direction (hereafter transverse velocity) ${\vl}^{\rm{obs}}\equiv\kappa d\mu_{\ell*}$,
where $d$ is the distance, $\kappa=4.74047$ is the constant for the change of $\kpc\,\mas\,\yr^{-1}$ to $\kms$, and $\mu_{\ell*}\equiv\mu_{\ell}\cos(b)$ is the proper motion in Galactic longitude. The median error in $\vl^{\rm{obs}}$ for the primary sample is $6\kms$ and $75\%$ of the stars have errors smaller than $10\kms$. This velocity can be corrected for the solar motion with: 
      \be\label{e_d1}
    \vl=\vl^{obs}-\Us\sin \ell+\Vs\cos \ell
   \ee  
\section{The method}\label{method}
Our procedure consists of comparing the median velocity of symmetric Galactic longitudes, that is $\ell$ and $-\ell$, in bins of longitude and distance on the Galactic plane ($\ell$, $d\cos(b)$). That is:
         \be\label{e_d2}
         \begin{split}
    \vl(+)-\vl(-)\equiv
    \widetilde{\kappa d\mu_{\ell*}}\,(\ell>0)-\widetilde{\kappa d\mu_{\ell*}}\,(\ell<0)-2\Us\sin |\ell|,
             \end{split}
             \ee
where the $\Vs\cos \ell$ term from \eqn{e_d1} cancels out. 
    In an axisymmetric Galaxy, $\vl$ is symmetric in $\ell$ and, therefore, we expect that $\vl(+)-\vl(-)=0$.  
Non-null values of $\vl(+)-\vl(-)$ show the  contribution of the non-axisymmetries. Note that asymmetries in the Galactic radial and azimuthal velocities may both contribute. This methodology has the advantage of being model-independent. Only an assumption on $\Us$ is required. We assume $\Us=9\kms$ (similar to determinations of  \citealt{Coskunoglu12011,Pasetto2012}, but see discussion in \Sec{strlim}). Asymmetries in $\vb^{\rm{obs}}\equiv\kappa d\mu_{b}$ are postponed to a forthcoming paper.

We estimate the ${95}\%$ confidence band of $\vl(+)-\vl(-)$ 
with  bootstrapping. We only consider bins with at least 5 stars (thus 10 stars in the pair $\ell>0$ and $\ell<0$). The typical dispersion of the bootstrapped median $\widetilde{\kappa d\mu_{\ell*}}$ (indicative of the precision of the median) is of 0.3 and $0.9\kms$ at a distance of 0.5 and $1\kpc$, respectively. This is much smaller than the individual stellar errors (\Sec{data}) due to the large number of stars in each bin.

\section{Tangential velocity asymmetry}\label{results}

 \begin{figure}
   \centering
   \includegraphics[width=0.35\textwidth]{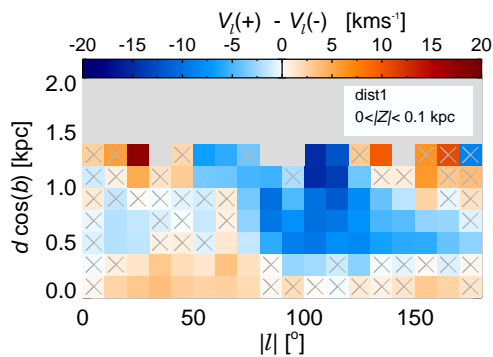}
      \caption{Difference between the median transverse velocity in Galactic longitude as a function of longitude and distance for symmetric Galactic longitudes $\ell>0$ and $\ell<0$. We use bins of $\Delta d \cos(b)={0.2}\kpc$ and $\Delta \ell=10\deg$. We plot a grey cross in bins where $\vl(+)-\vl(-)$ is  statistically consistent with 0 with a ${95}\%$ confidence, i.e. where the observations are compatible with an axisymmetric Galaxy. The grey region shows bins {with insufficient} data (we require at least 5 stars in each $\ell>0$ and $\ell<0$ bin). We have fixed the colour scale to $20\kms$.}
         \label{figdist1}
   \end{figure}
   
   In \fig{figdist1} we plot\footnote{Note that this is the same quantity plotted in figure~6 of A16, which was called $\D-\De$ there. Note also that in A16 we used larger bins in distance and a different vertical range in the plots.} $\vl(+)-\vl(-)$ as a function of longitude and distance on the plane for the primary sample (\Sec{data}) using dist1 (see \Sec{strlim} for the consistency with other distance estimates). Bins marked with crosses have values compatible with 0 given their ${95}\%$ confidence band. For an axisymmetric Galaxy, one would expect values compatible with 0 everywhere in this plot. But we see that {${57}\%$ of the bins probed} present an unbalance tangential motion that is statistically significant {at {95}\% level}, i.e. positive or negative $\vl(+)-\vl(-)$. {If the distribution of errors in $\vl(+)-\vl(-)$ was Gaussian and the Galaxy was axisymmetric, only  5\% of bins would deviate from 0 at this level of confidence, thus much smaller than the measured fraction.} 

{An alternative test to the hypothesis of axisymmetry is to assume that the distribution of $\vl(+)-\vl(-)$ for an axisymmetric Galaxy is a Gaussian centred at 0 with a sigma derived from the actual data. We did 500 bootstraps of the median $\vl$ at each bin, obtaining effectively 124750 bootstrapped $\vl(+)-\vl(-)$ (combinations without repetition) for each pair of bins, from which we computed $\sigma$. Given the measured $\vl(+)-\vl(-)$ and the assumed distribution under the null hypothesis, we find that  71$\%$ of bins deviate from axisymmetry at 95\% level (p<0.05). 
 By combining all the individual p values with the Fisher's method, we infer that the hypothesis of axisymmetry can be rejected with a 
p value < 0.001.
}

   The asymmetry is such that the median velocity difference oscillates between {-15} and $18\kms$, has a median (of absolute values) of ${2.6\kms}$ and a median absolute deviation ({MAD}) of ${2.7\kms}$ (considering {all bins}). 
   {The median asymmetry is minimum ($1.4\kms$) for the distance between 0.2 and 0.4 $\kpc$, and maximum (${8.8\kms}$) at a distance between 1.2 and $1.4\kpc$.  It is minimum (${1.2\kms}$) at longitudes of ${\pm5\deg}$ and maximum (${15\kms}$) at longitudes of ${\pm105\deg}$.}
   A percentage of ${47}\%$ of all bins probed show absolute differences larger than $2\kms$ that are statistically significant {at 95\% level}, $24\%$ larger than $5\kms$, and ${7\%}$ larger than $10\kms$. {All} these numbers indicate that the fraction of the solar suburbs with kinematics showing signs of non-axisymmetry is quite large.

The differences in transverse velocity of \fig{figdist1}  show a pattern of a scale of several tens of degrees and about $0.5\kpc$ in distance. We observe  a large region of negative differences (blue colours) in the range $|\ell|\sim [70,180]\deg$. The positive differences (red colours) are mostly located at close distances, and in the inner and outer disk directions at the farthest distances probed.

\begin{figure}
   \centering
   \includegraphics[width=0.25\textwidth]{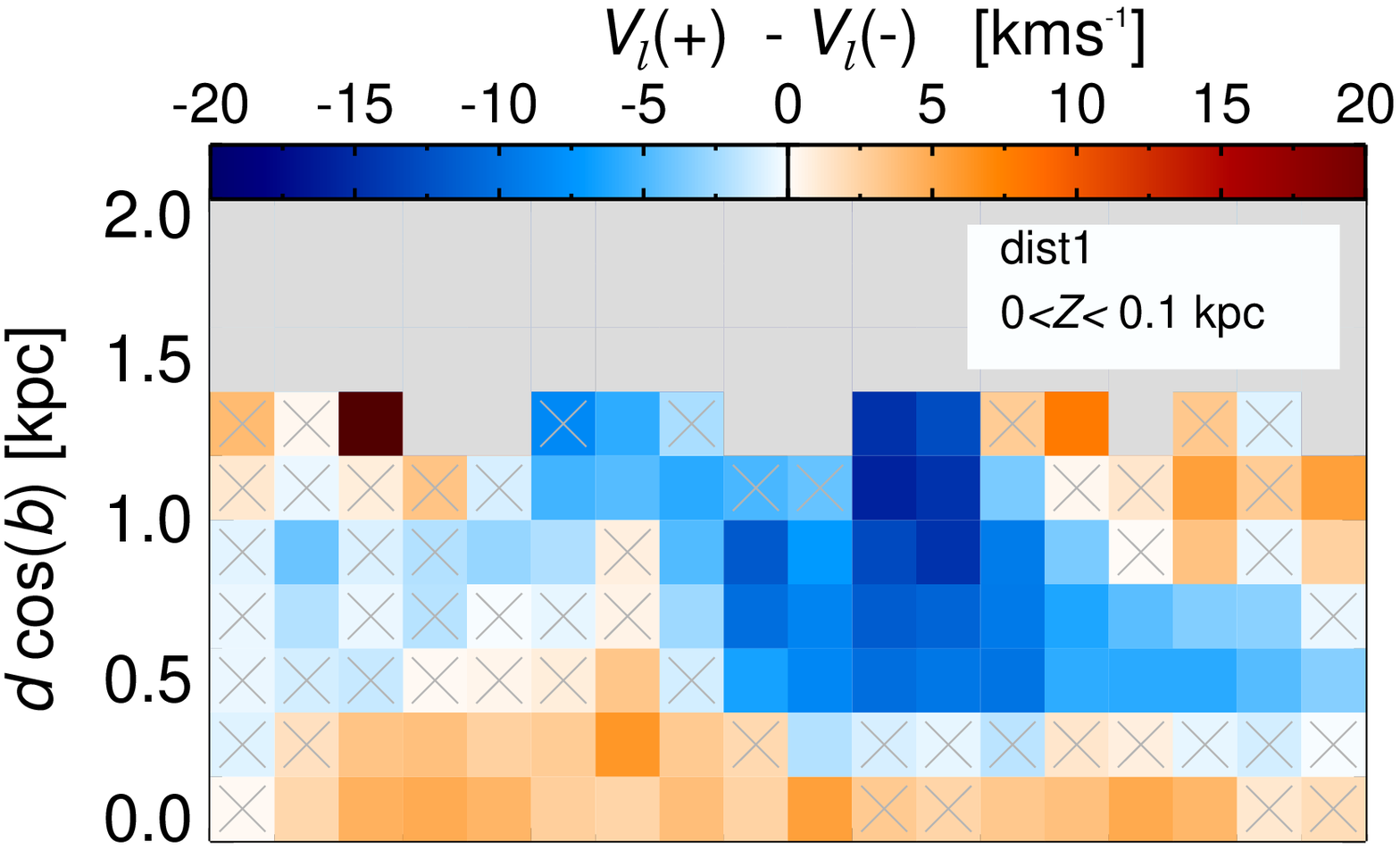}\hspace{-0.6cm}   
   \includegraphics[width=0.25\textwidth]{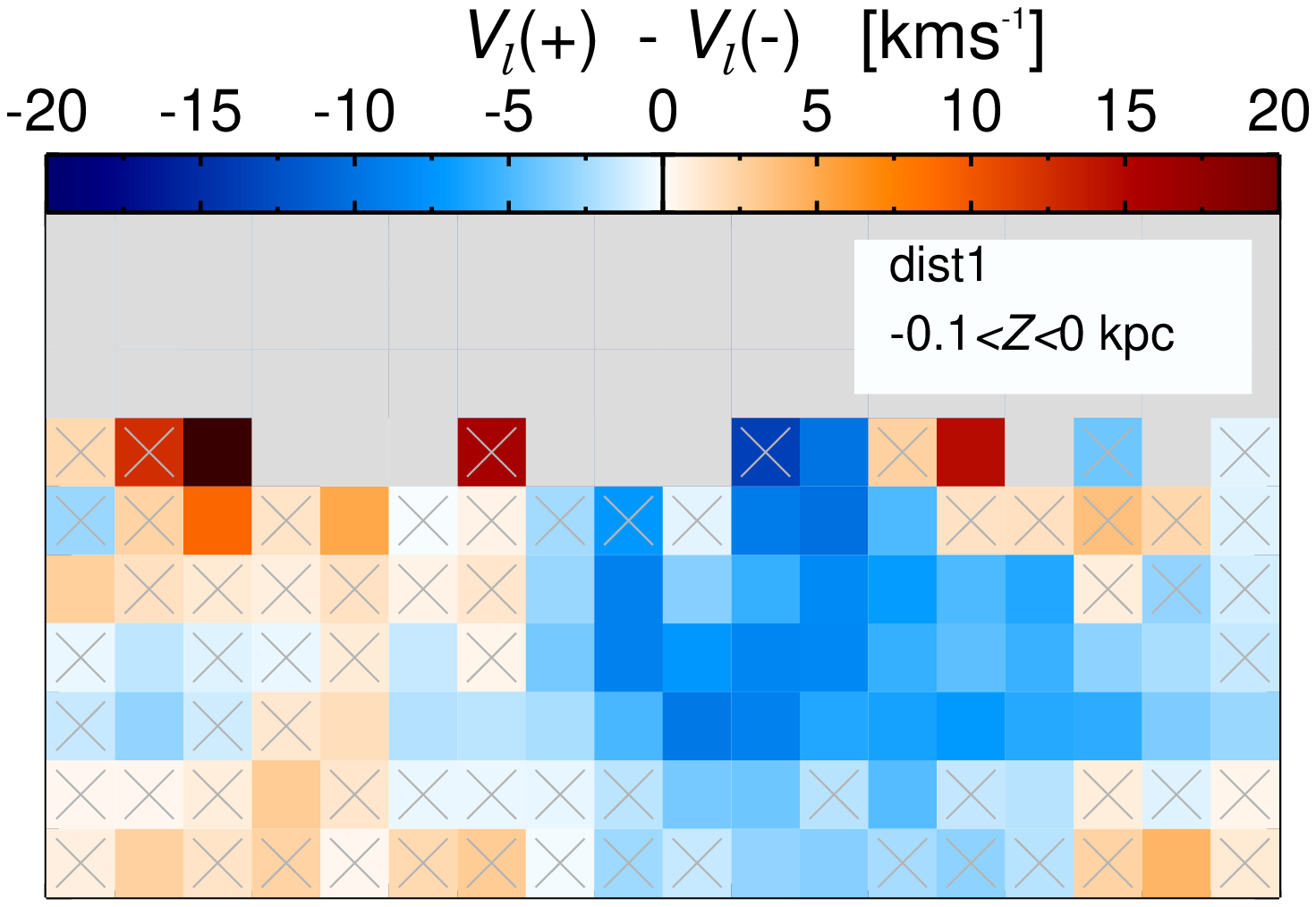}
\vspace{-0.6cm}   

   \includegraphics[width=0.25\textwidth]{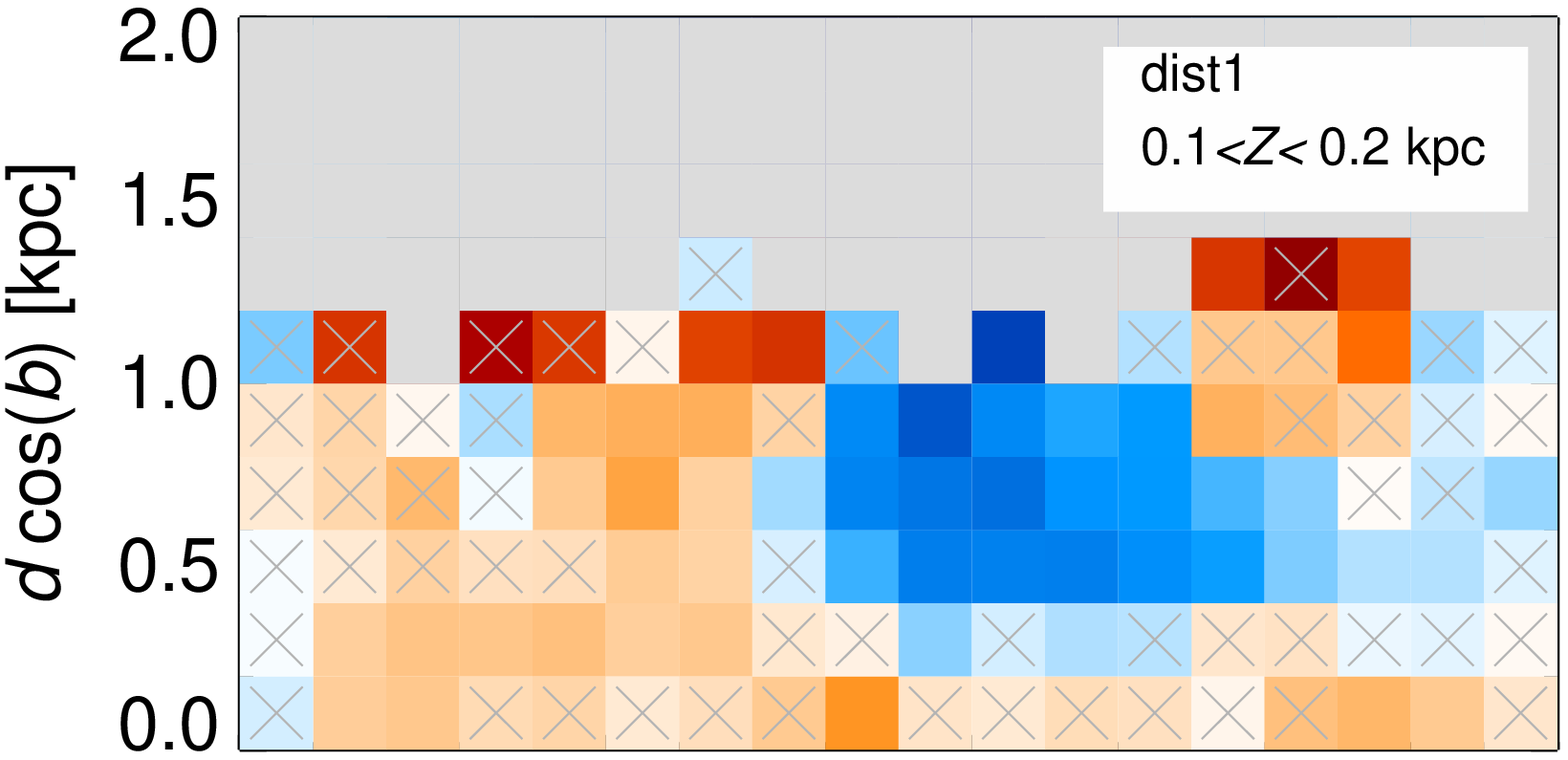}\hspace{-0.6cm}   
   \includegraphics[width=0.25\textwidth]{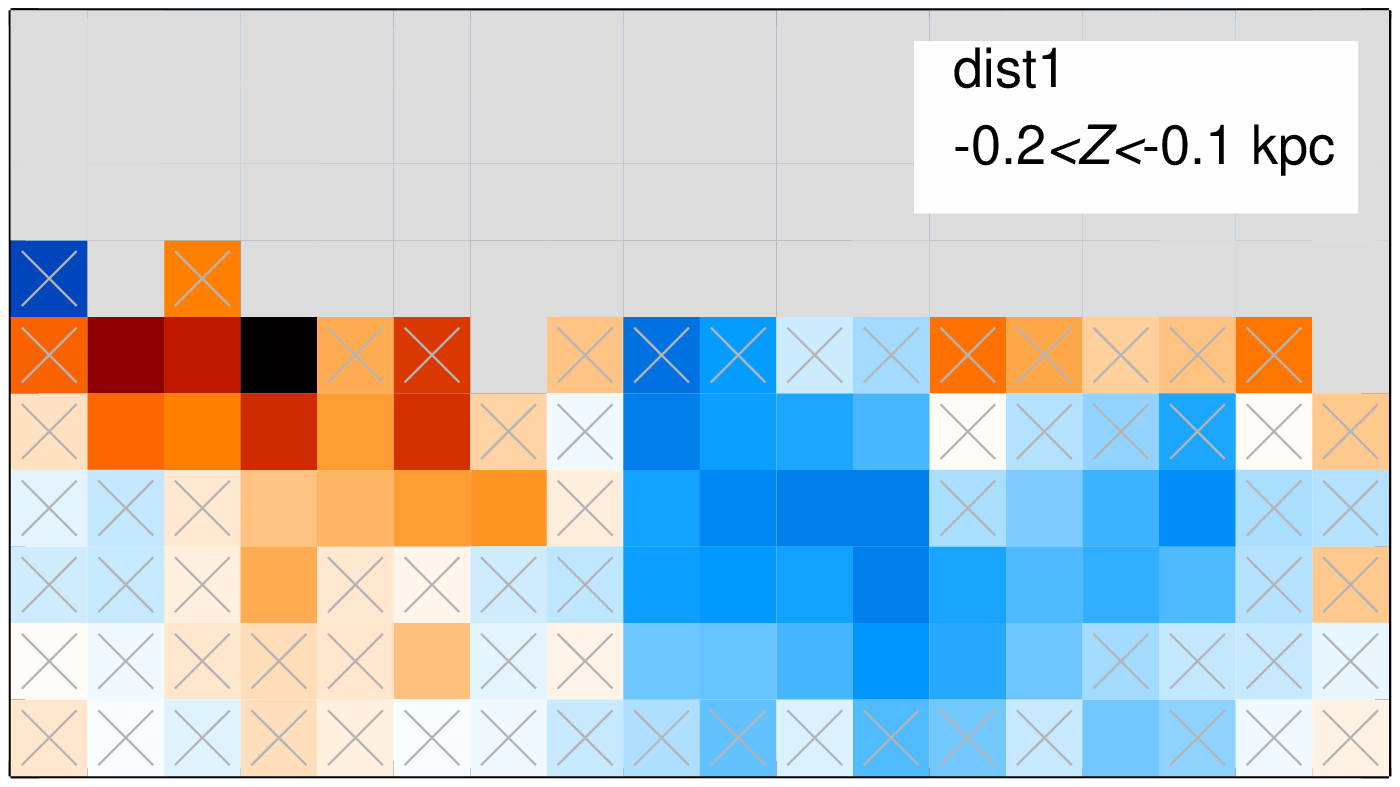}
\vspace{-0.6cm}   

    \includegraphics[width=0.25\textwidth]{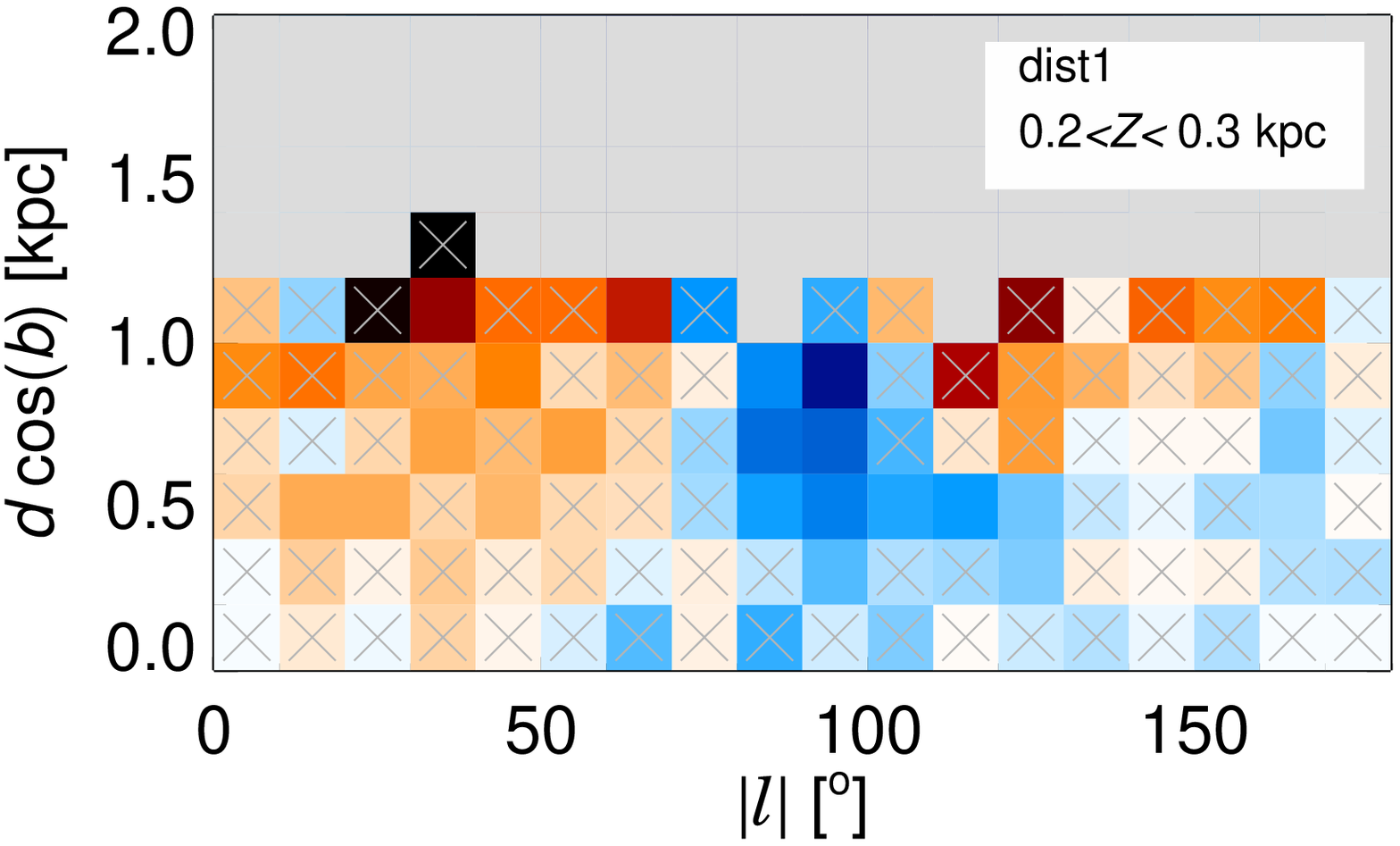}  \hspace{-0.66cm}    
   \includegraphics[width=0.25\textwidth]{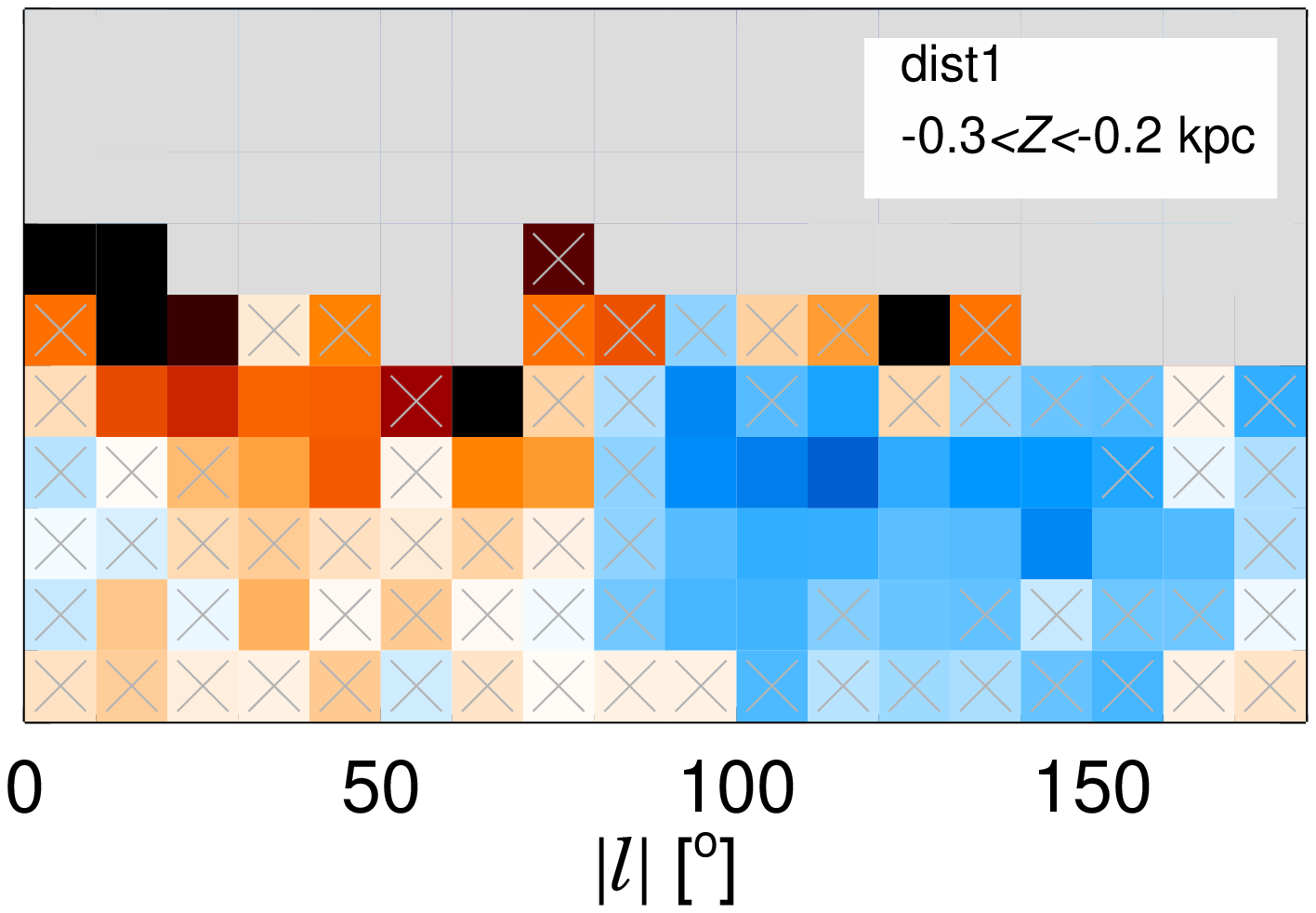}   

     \caption{Same as \fig{figdist1} but in different layers in $Z$.}
         \label{figZ}
   \end{figure}

     \Fig{figZ} shows the velocity asymmetries for different layers in $Z$ as indicated in the legends. The asymmetry extends at least up and down in the plane to $|Z|=300\pc$. Beyond this height it looses significance due to the lack of data. We find a mild dependence in $Z$. The region with negative asymmetry for $|\ell|\gtrsim70\deg$ is present at all $Z$. But there is a large region with positive asymmetry at $|\ell|\lesssim70\deg$ only present at the higher $Z$ probed. Beyond $|Z|>100\pc$ the asymmetry seems to be dominated by a duality of positive and negative asymmetry, while most of the small scale pattern is significant only at low $Z$. We do not find differences with small changes in the $Z\sun$ assumed.

       \begin{figure}
   \centering
      \includegraphics[width=0.245\textwidth]{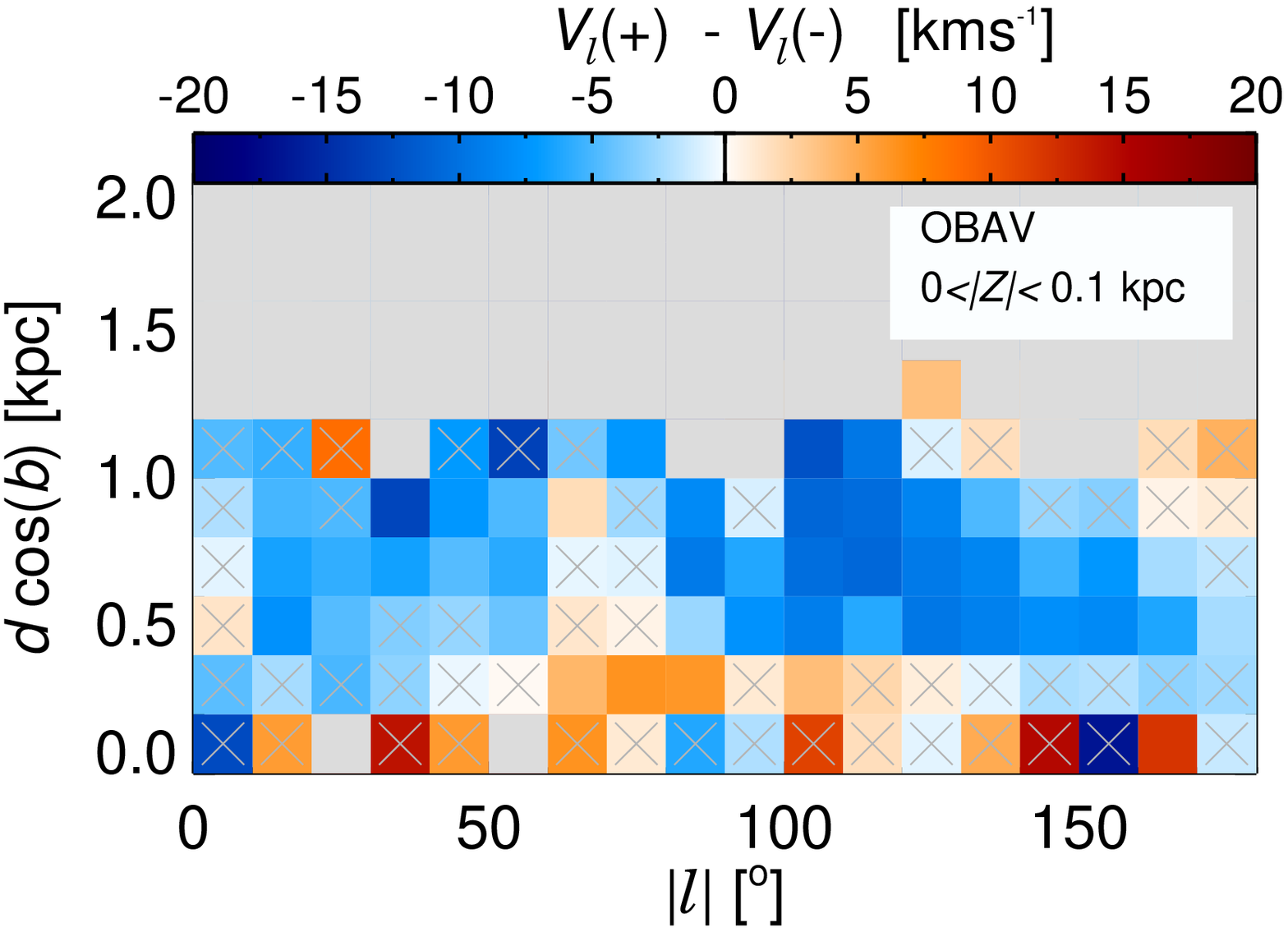}\hspace{-0.6cm}       
      \includegraphics[width=0.245\textwidth]{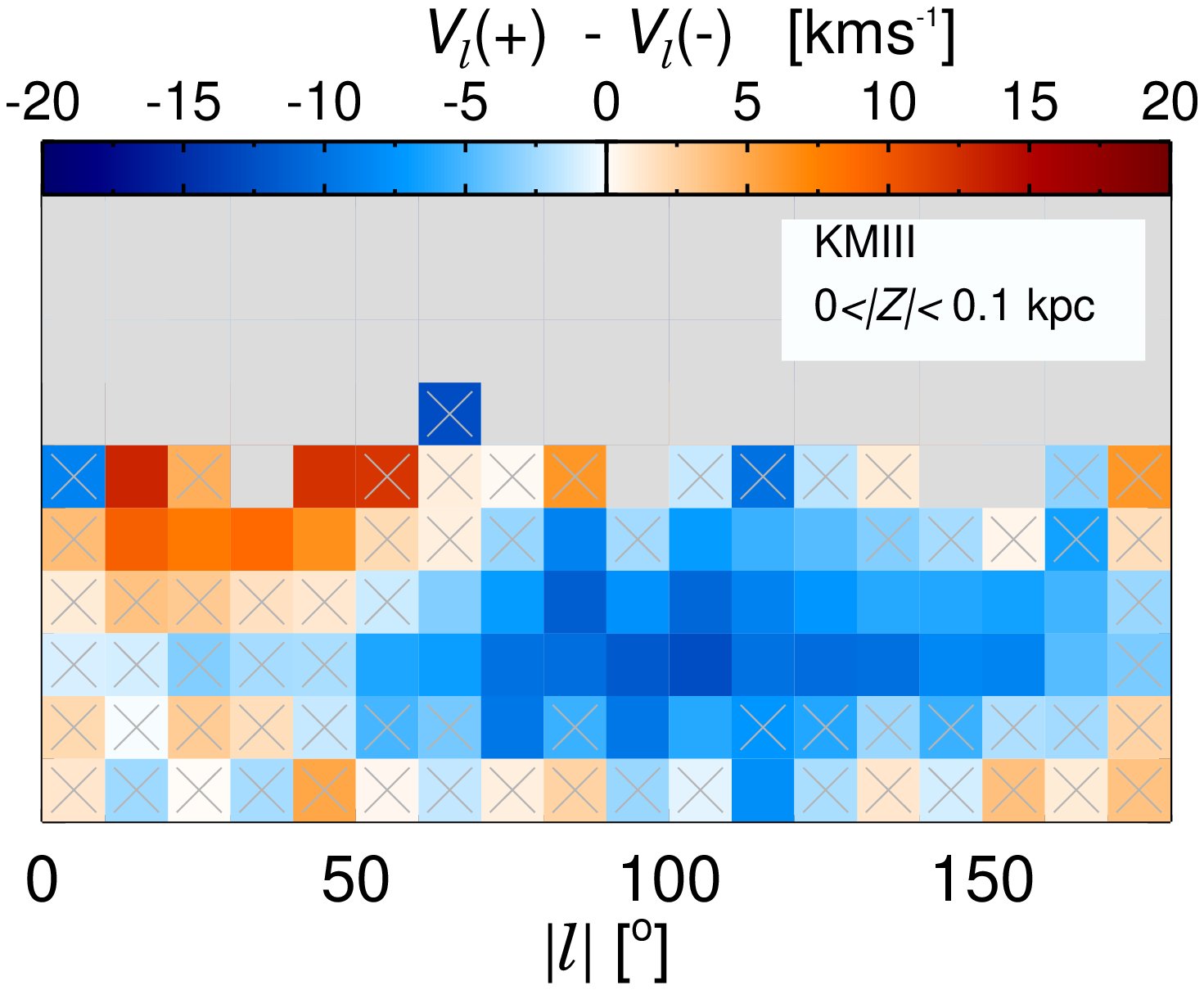}  
        \caption{Same as \fig{figdist1} but for  different {stellar types}. {The plot for FGKV is omitted since it looks essentially as \fig{figdist1}.}}
         \label{figST}
   \end{figure}

\Fig{figST} shows the velocity asymmetries for the different stellar groups described in \Sec{data} (except for the  FGKV since it looks essentially as \fig{figdist1}.). Interestingly, the large blue region of negative differences appears for all sub-samples but the asymmetry differs substantially in the inner Galaxy ($\ell\lesssim70$): it is negative for the OBAV group, it is both positive and negative for the FGKMV group, and it is mostly positive for the KMIII group at the furthest distances (there is not enough statistics at closer distances for this group). The median velocity asymmetry (for statistically significant bins) is largest for the KMIII group ($8\kms$), followed by the OBAV ($7\kms$) and by the FGKMV (${5}\kms$). 


   \section{Robustness tests}\label{strlim}
  
    \begin{figure}
   \centering
   \includegraphics[width=0.245\textwidth]{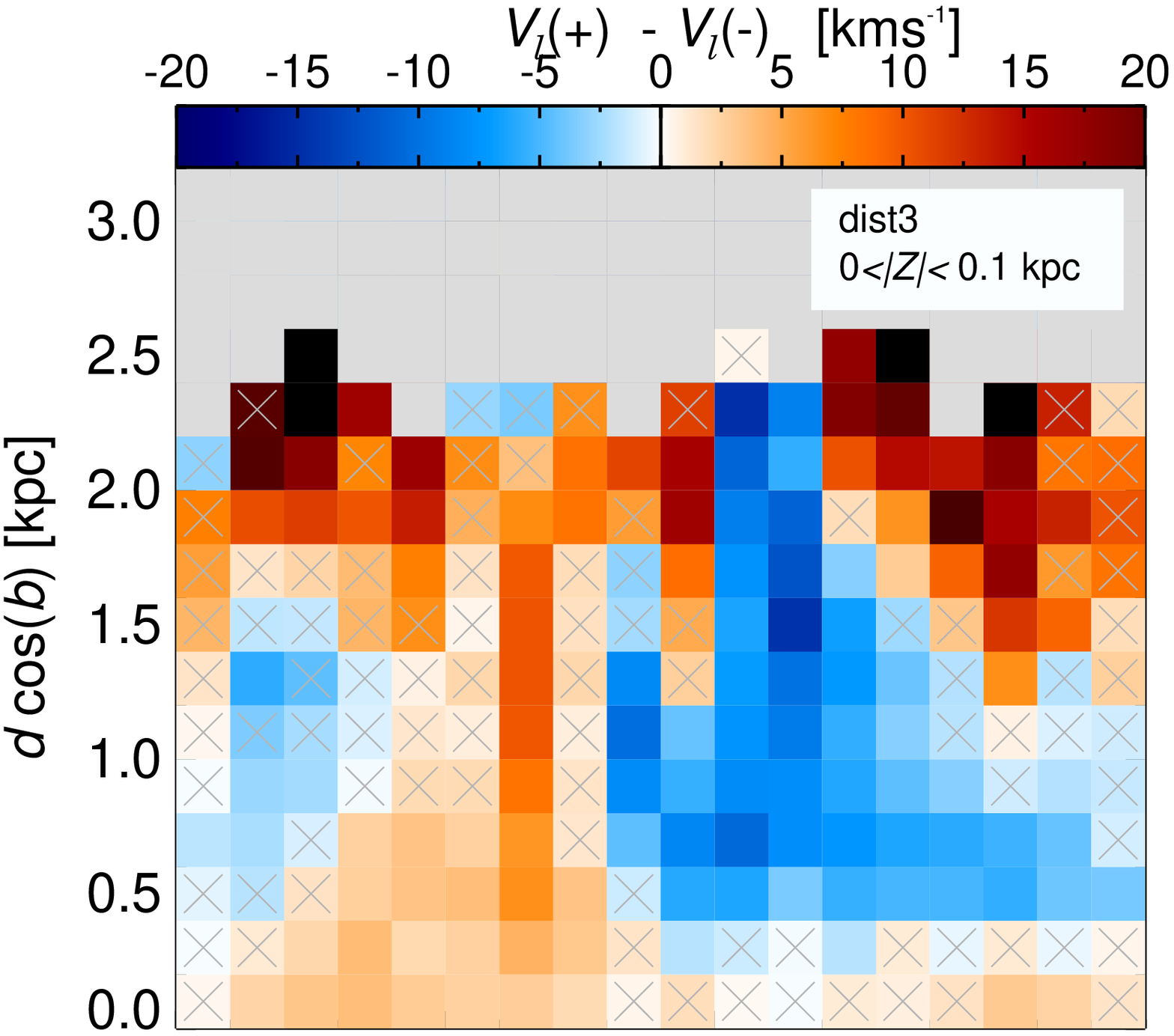}\hspace{-0.6cm}   
       \includegraphics[width=0.245\textwidth]{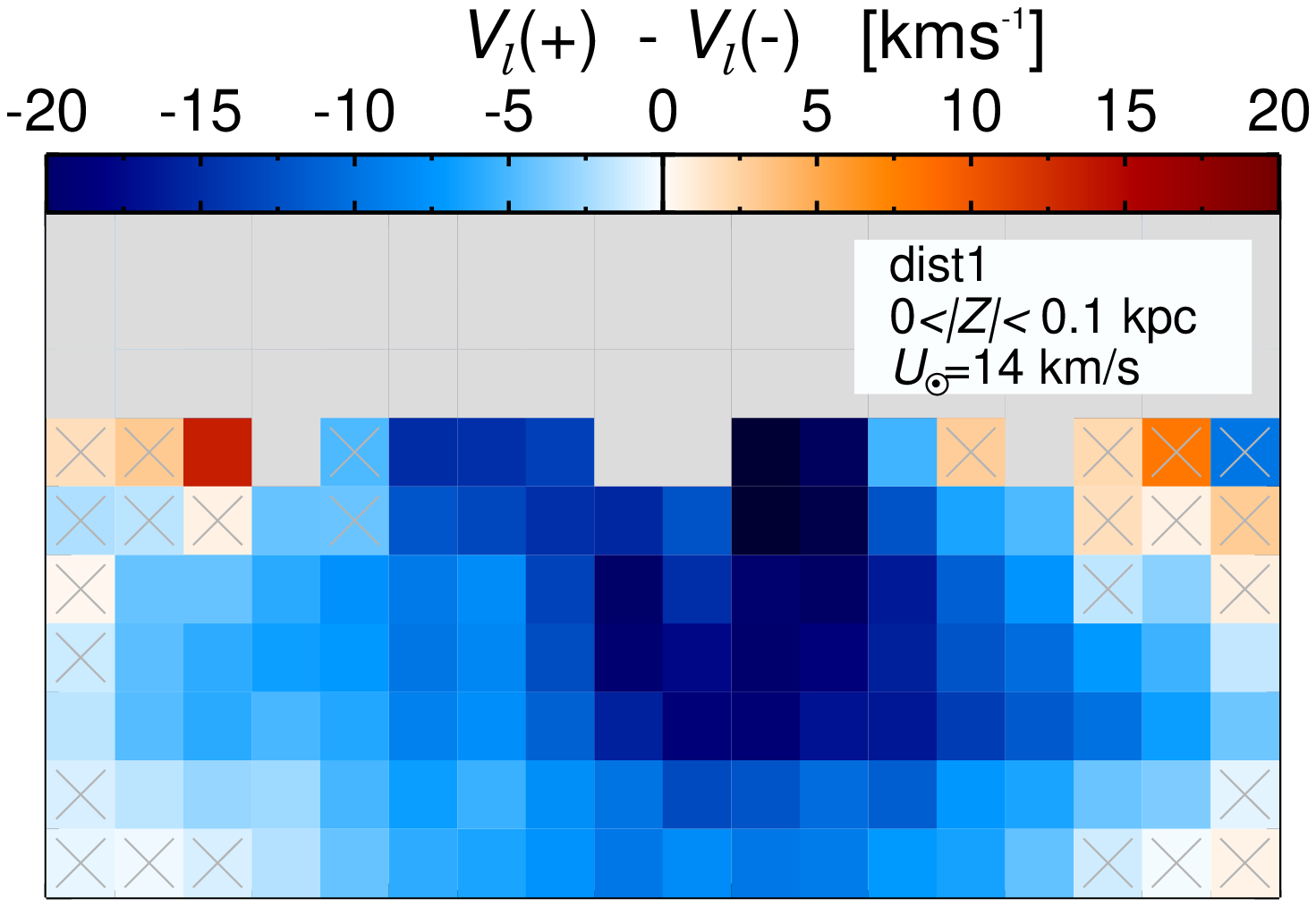}      
       \vspace{-0.6cm} 
       
    \includegraphics[width=0.245\textwidth]{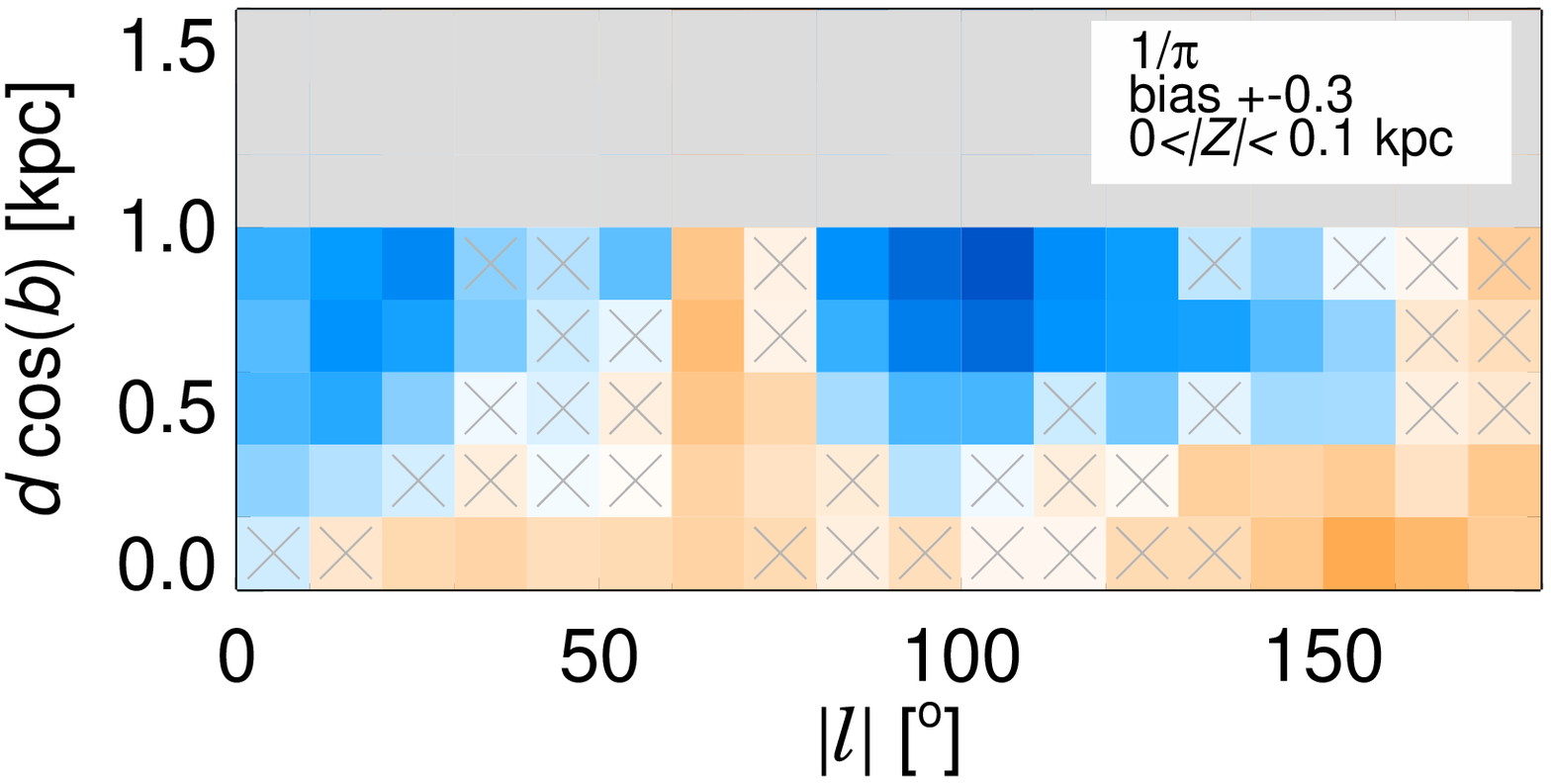}\hspace{-0.6cm} 
   \includegraphics[width=0.245\textwidth]{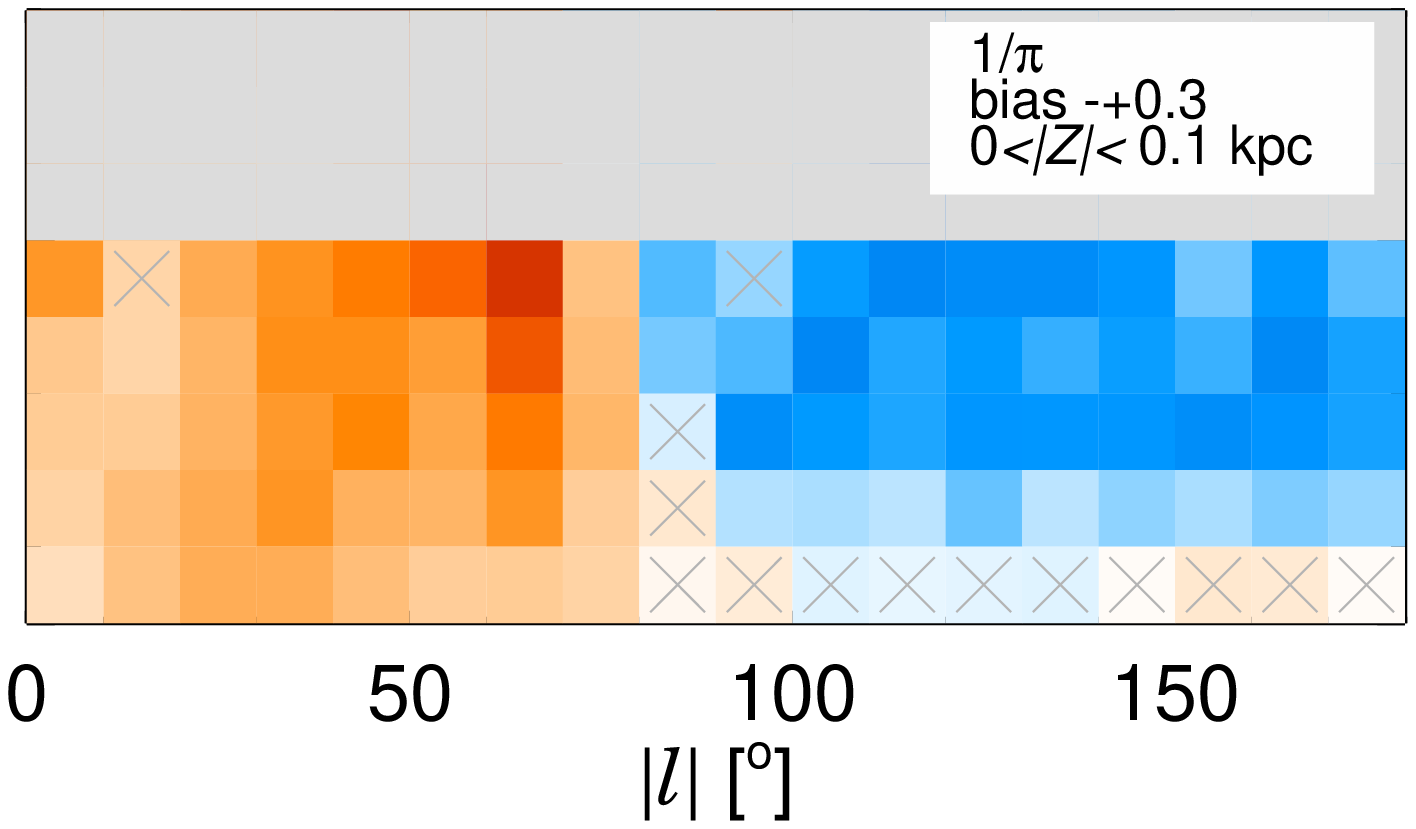}
      \caption{Same as \fig{figdist1} but for dist3 ({\em first {top} panel}), assuming $\Us=14\kms$ ({\em {second top} panel}), correcting for an assumed systematic bias in the parallax of $+0.3\mas$ for $\ell>0$ and of $-0.3\mas$ for $\ell<0$ ({\em {first bottom} panel}), and the reverse bias ({\em {second bottom} panel}). {The bottom panels are cut at $1\kpc$ where the error in distance starts to be larger than the bin size.} }
         \label{figdist3}
   \end{figure}

Here we perform tests to show the robustness of our results and check that the limitations of the TGAS data and of our method do not contribute to induce/increase the observed velocity asymmetry between $\ell>0$ and $\ell<0$.
   
   {\it General astrometric quality.} When we select stars in the first quartile with better astrometric quality ({\it astrometric\_excess\_noise}$<0.37$) we observe no significant differences with \fig{figdist1}. The values of $\vl(+)-\vl(-)$ differ in median only by ${1.4}\kms$. 
 {Also, the sign of the asymmetry is the same in all bins in common where $\vl(+)-\vl(-)$ is significant (40 bins).}
   
   {\it Distance estimation choice.} We see that the main difference when using different distance estimations is in the overall distance scale which extends much further for $1/\pi$, dist2 and dist3 (\fig{figdist3}, first panel) compared to dist1. Additionally, for these alternative distance estimates, the  magnitude of the asymmetry in velocity increases as a function of distance suspiciously for all longitudes, reaching extreme values (black colours in the first panel of \fig{figdist3}). For instance, with dist2 $\vl(+)-\vl(-)$ has a median of ${6.3}\kms$ and {24}$\%$ of the bins show differences larger than $10\kms$. These large asymmetries could be due to an overestimation of the distance which leads to an overestimation of $\vl$. This effect is only mildly observed for dist1 (but note the same effect at high $Z$ in \fig{figZ}), hence, our preferred choice of this distance estimate. 
   Apart from these differences, our main results do not depend significantly on the distance estimate: the sign of the pattern is the same in {90}$\%$, {81}$\%$ and {91}$\%$ of the significant bins when comparing dist1 with $1/\pi$, dist2 and dist3, respectively.
   
  {\it Parallax accuracy.} As recommended in \citet{Arenou2017}, a possible parallax bias of $\sim0.3\mas$ that could be non-uniform in the sky has to be considered in the analysis of TGAS data. Here we see that the distances with this systematic error taken into account from \citet{Astraatmadja2016} slightly shorten the distance scale but do not change the asymmetry. We have also repeated our calculations using $1/\pi$ and correcting for a systematic bias, that is effectively adding or subtracting $0.3\mas$ to the parallax. We show here two cases where we add $-0.3\mas$ for $\ell>0$ and  $+0.3\mas$ for $\ell<0$, and the reverse (\fig{figdist3}, {bottom} panels). Note that these are the worse case scenarios {with} the bias {contributing}  maximally to the velocity asymmetry in \eqn{e_d2}. In these cases the asymmetry pattern changes slightly, especially in the directions to the inner and outer Galaxy, but is overall preserved. 
  We conclude that the presumed bias in parallax can not be responsible for the global velocity asymmetry observed. {Note also that the negative sign of the asymmetry at $l\sim100\deg$ does not change even under the more extreme systematics considered here. This is, therefore, a very robust result that models of the asymmetry must reproduce. } 

{\it Correlation between parallax and proper motion.}  The astrometric correlations can be of up to $\pm1$  in certain sky regions in the TGAS data (see figure C.1 in \citealt{Arenou2017}). However, the velocity asymmetry is not induced by these correlations. We have tested this by, first,  adding uncorrelated noise equal to 3 times the standard errors reported in the catalogue, which is enough to break the correlations. We only observe little changes, that might well be due to introducing larger errors and not to the correlations, but the overall asymmetry pattern is preserved. Secondly, we have added extra correlated noise using the individual reported standard errors and setting all astrometric correlations $i,j$ to $\rho_{ij}=\pm1$ (keeping its original sign). This does not increase the observed velocity asymmetry.  

{\it Sky coverage, completeness, extinction.} The TGAS catalogue is incomplete  and has a non-uniform sky coverage (e.g., see figure 5 of \citealt{Arenou2017}). Due to this and to extinction, there are differences in the number of observed stars in the symmetric bins $(\ell,d\cos(b))$ and $(-\ell,d\cos(b))$. However, these will not bias the median transverse velocity but only change its precision, assuming there are no additional selection effects. 
{Besides, if in the pairs of bins there was a difference in the vertical distribution of stars (for instance a different median $Z$  at positive and negative longitudes), the measured asymmetry could be due to a different $\vl$ as a function of $Z$, which is expected in an axisymmetric galaxy. However, the velocity changes with $Z$ in the thin layer that we select ($200 \pc$) are much smaller than the velocity asymmetry that we measure:  taking equation 13 of \citet{Bond2010}, the velocity at $Z=0$ would change only by $1\kms$ at $Z=100\pc$.}


{\it Assumption of $\Us$.}  Ideally, one should fit the value of $\Us$ at the same time of a non-axisymmetry model to the observed velocity asymmetry. Here our analysis requires an assumption for $\Us$ (\eqn{e_d2}).
However, note that a different $\Us$ can not smooth out completely the asymmetry at all Galactic longitudes because the term $-2\Us\sin|\ell|$ has always the same sign. It will only modify the pattern and sign of the asymmetry. 
E.g. , if we use $\Us=14\kms$ \citep{Schonrich2012}, the asymmetry becomes negative everywhere (\fig{figdist3} {top} right panel) reaching values down to $-{25}\kms$ and with ${34}\%$ of the bins with asymmetries as large as $10\kms$, but does not disappear.
On the other hand, one can estimate a value of $\Us$ from the data by averaging  the quantity $\Us= \frac{\widetilde{\kappa d\mu_{\ell*}}\,(\ell>0)-\widetilde{\kappa d\mu_{\ell*}}\,(\ell<0)}{2\sin |\ell|}$ for all bins (i.e. supposing that $\vl(+)-\vl(-)=0$, in other words, that there is no net contribution from non-axisymmetries). This value is $\Us={8.3\pm0.6}\kms$, and thus, for our choice of $\Us=9\kms$ the total net asymmetry is the smallest one compared to other values.

To conclude, 
with the information currently available to us on the quality and limitations of the TGAS data and of our method, the measurement of the velocity asymmetry is robust.

%
%

\section{Discussion and conclusions}\label{conclusion}

We have detected velocity asymmetries when comparing the median transverse velocity in Galactic longitude for positive and negative longitudes using the Gaia-TGAS catalogue following a model-independent approach. The sign of the velocity differences follows a pattern that depends on the distance and direction.
The  velocity asymmetry reaches values larger than $10\kms$ for ${7}\%$ of the region studied. 
This asymmetry, which extends to all distances and directions probed, indicates that the stellar motion in the disk is highly non-axisymmetric.

Part of the asymmetry (in the direction of the outer disk) is present for all the stellar types considered here. This points towards a common dynamical origin of the asymmetry.  The differences seen 
for the young sample (not yet phase-mixed) can be due to imprints of the velocities at birth or structures such as the Gould's belt \citep{Lesh1968,Comeron1994}. The differences when comparing dwarfs with giant stars could be due to the same perturbation acting different on different mean ages.

Regarding the magnitude of the asymmetry, the values that we find are similar to previous determinations of streaming motion. For example, star-forming regions deviate from rotation typically by $10\kms$ \citep{Reid2014,Honma2012,Rygl2012}. In external galaxies, radial streaming motions of $7\kms$ are observed \citep{Rix1995}. For the Milky Way, velocity gradients in the radial direction are of $3\kmskpc$ \citep{Siebert2011a} and fluctuations with amplitude of $10\kms$ have been measured \citep{Bovy2015}. 

In A16, we studied the asymmetries in transverse velocity for a series of disc simulations with spiral structure. Some models followed the Tight-Winding-Approximation (TWA) and some were N-Body models. The magnitude of the typical velocity asymmetries of the models were of the order of $\sim2\kms$ but up to $10\kms$, thus, resembling those found here (see also \citealt{Faure2014,Grand2016} for alternative but similar predictions). However, the spatial scales of the variations of the asymmetry patterns were larger in distance compared to the data, except for the model of transient arms.
Also the particular pattern of the observed asymmetry and, in particular, its sign does not follow  what we saw in the vast majority of models that were built to resemble the spiral structure of the Galaxy (see figure 6 of A16 but note the different range of distance).
 However, a quantitative fit of the model exploring the whole range of spiral parameters is necessary to draw definitive conclusions (Antoja, Roca-Fabrega et al. in preparation).

This asymmetry could also be attributed to  
the Galactic bar that can deviate the velocities from axisymmetry by about $5$-$10\kms$ near the Sun \citep{Monari2014,Bovy2015}, thus compatible with the data here. A perturbation from a satellite could excite breathing or other disk modes \citep{Gomez2013,Widrow2014} but little attention is put on its effects on the in-plane velocities.  An elliptic potential induced by a non-spherical halo can also perturb the in-plane velocities \citep{Kuijken1994}.
 How well these other models reproduce the observed asymmetry needs to be investigated. 
Several agents may contribute simultaneously to it, creating a quite intricate Galaxy disk. We hope to decipher it with the advent of new data (Gaia DR2 and follow-up surveys) and models that combine internal and external agents driving the evolution of the Milky Way disk.

\begin{acknowledgements}
{We thank the referee for the careful
reading and advice.}
This work was supported by the MINECO (Spanish Ministry of Economy) - FEDER through grant ESP2016-80079-C2-1-R and ESP2014-55996-C2-1-R and MDM-2014-0369 of ICCUB (Unidad de Excelencia 'Maria de Maeztu'). This work has made use of data from the European Space Agency (ESA)
mission {\it Gaia} (\url{https://www.cosmos.esa.int/gaia}), processed by
the {\it Gaia} Data Processing and Analysis Consortium (DPAC,
\url{https://www.cosmos.esa.int/web/gaia/dpac/consortium}). Funding
for the DPAC has been provided by national institutions, in particular
the institutions participating in the {\it Gaia} Multilateral Agreement. We also thank the Gaia Project Scientist Support Team, DPAC and Anthony Brown for the PyGaia code that has been used in this research.
\end{acknowledgements}

\vspace{-0.7cm}
\bibliographystyle{aa} 
\bibliography{mybib}

\begin{thebibliography}{30}
\expandafter\ifx\csname natexlab\endcsname\relax\def\natexlab#1{#1}\fi

\bibitem[{{Antoja} {et~al.}(2016){Antoja}, {Roca-F{\`a}brega}, {de Bruijne}, \&
  {Prusti}}]{Antoja2016}
{Antoja}, T., {Roca-F{\`a}brega}, S., {de Bruijne}, J., \& {Prusti}, T. 2016,
  \aap, 589, A13

\bibitem[{{Arenou} {et~al.}(2017){Arenou}, {Luri}, {Babusiaux}, {Fabricius},
  {Helmi}, {Robin}, {Vallenari}, {Blanco-Cuaresma}, {Cantat-Gaudin},
  {Findeisen}, {Reyl{\'e}}, {Ruiz-Dern}, {Sordo}, {Turon}, {Walton}, {Shih},
  {Antiche}, {Barache}, {Barros}, {Breddels}, {Carrasco}, {Costigan},
  {Diakit{\'e}}, {Eyer}, {Figueras}, {Galluccio}, {Heu}, {Jordi},
  {Krone-Martins}, {Lallement}, {Lambert}, {Leclerc}, {Marrese}, {Moitinho},
  {Mor}, {Romero-G{\'o}mez}, {Sartoretti}, {Soria}, {Soubiran}, {Souchay},
  {Veljanoski}, {Ziaeepour}, {Giuffrida}, {Pancino}, \&
  {Bragaglia}}]{Arenou2017}
{Arenou}, F., {Luri}, X., {Babusiaux}, C., {et~al.} 2017, \aap, 599, A50

\bibitem[{{Astraatmadja} \& {Bailer-Jones}(2016)}]{Astraatmadja2016}
{Astraatmadja}, T.~L. \& {Bailer-Jones}, C.~A.~L. 2016, \apj, 833, 119

\bibitem[{{Bond} {et~al.}(2010){Bond}, {Ivezi{\'c}}, {Sesar}, {Juri{\'c}},
  {Munn}, {Kowalski}, {Loebman}, {Ro{\v s}kar}, {Beers}, {Dalcanton},
  {Rockosi}, {Yanny}, {Newberg}, {Allende Prieto}, {Wilhelm}, {Lee},
  {Sivarani}, {Majewski}, {Norris}, {Bailer-Jones}, {Re Fiorentin}, {Schlegel},
  {Uomoto}, {Lupton}, {Knapp}, {Gunn}, {Covey}, {Allyn Smith}, {Miknaitis},
  {Doi}, {Tanaka}, {Fukugita}, {Kent}, {Finkbeiner}, {Quinn}, {Hawley},
  {Anderson}, {Kiuchi}, {Chen}, {Bushong}, {Sohi}, {Haggard}, {Kimball},
  {McGurk}, {Barentine}, {Brewington}, {Harvanek}, {Kleinman}, {Krzesinski},
  {Long}, {Nitta}, {Snedden}, {Lee}, {Pier}, {Harris}, {Brinkmann}, \&
  {Schneider}}]{Bond2010}
{Bond}, N.~A., {Ivezi{\'c}}, {\v Z}., {Sesar}, B., {et~al.} 2010, \apj, 716, 1

\bibitem[{{Bovy}(2017)}]{Bovy2017}
{Bovy}, J. 2017, \mnras, 468, L63

\bibitem[{{Bovy} {et~al.}(2015){Bovy}, {Bird}, {Garc{\'{\i}}a P{\'e}rez},
  {Majewski}, {Nidever}, \& {Zasowski}}]{Bovy2015}
{Bovy}, J., {Bird}, J.~C., {Garc{\'{\i}}a P{\'e}rez}, A.~E., {et~al.} 2015,
  \apj, 800, 83

\bibitem[{{Carlin} {et~al.}(2013){Carlin}, {DeLaunay}, {Newberg}, {Deng},
  {Gole}, {Grabowski}, {Jin}, {Liu}, {Liu}, {Luo}, {Yuan}, {Zhang}, {Zhao}, \&
  {Zhao}}]{Carlin2013}
{Carlin}, J.~L., {DeLaunay}, J., {Newberg}, H.~J., {et~al.} 2013, \apjl, 777,
  L5

\bibitem[{{Chen} {et~al.}(2001){Chen}, {Stoughton}, {Smith}, {Uomoto}, {Pier},
  {Yanny}, {Ivezi{\'c}}, {York}, {Anderson}, {Annis}, {Brinkmann}, {Csabai},
  {Fukugita}, {Hindsley}, {Lupton}, {Munn}, \& {SDSS Collaboration}}]{Chen2001}
{Chen}, B., {Stoughton}, C., {Smith}, J.~A., {et~al.} 2001, \apj, 553, 184

\bibitem[{{Co{\c s}kuno{\v g}lu} {et~al.}(2011){Co{\c s}kuno{\v g}lu}, {Ak},
  {Bilir}, {Karaali}, {Yaz}, {Gilmore}, {Seabroke}, {Bienaym{\'e}},
  {Bland-Hawthorn}, {Campbell}, {Freeman}, {Gibson}, {Grebel}, {Munari},
  {Navarro}, {Parker}, {Siebert}, {Siviero}, {Steinmetz}, {Watson}, {Wyse}, \&
  {Zwitter}}]{Coskunoglu12011}
{Co{\c s}kuno{\v g}lu}, B., {Ak}, S., {Bilir}, S., {et~al.} 2011, \mnras, 412,
  1237

\bibitem[{{Comeron} \& {Torra}(1994)}]{Comeron1994}
{Comeron}, F. \& {Torra}, J. 1994, \aap, 281, 35

\bibitem[{{Faure} {et~al.}(2014){Faure}, {Siebert}, \& {Famaey}}]{Faure2014}
{Faure}, C., {Siebert}, A., \& {Famaey}, B. 2014, \mnras, 440, 2564

\bibitem[{{Gaia Collaboration} {et~al.}(2016{\natexlab{a}}){Gaia
  Collaboration}, {Brown}, {Vallenari}, {Prusti}, {de Bruijne}, {Mignard},
  {Drimmel}, {Babusiaux}, {Bailer-Jones}, {Bastian}, \&
  et~al.}]{GaiaCollaboration2016b}
{Gaia Collaboration}, {Brown}, A.~G.~A., {Vallenari}, A., {et~al.}
  2016{\natexlab{a}}, \aap, 595, A2

\bibitem[{{Gaia Collaboration} {et~al.}(2016{\natexlab{b}}){Gaia
  Collaboration}, {Prusti}, {de Bruijne}, {Brown}, {Vallenari}, {Babusiaux},
  {Bailer-Jones}, {Bastian}, {Biermann}, {Evans}, \&
  et~al.}]{GaiaCollaboration2016}
{Gaia Collaboration}, {Prusti}, T., {de Bruijne}, J.~H.~J., {et~al.}
  2016{\natexlab{b}}, \aap, 595, A1

\bibitem[{{G{\'o}mez} {et~al.}(2013){G{\'o}mez}, {Minchev}, {O'Shea}, {Beers},
  {Bullock}, \& {Purcell}}]{Gomez2013}
{G{\'o}mez}, F.~A., {Minchev}, I., {O'Shea}, B.~W., {et~al.} 2013, \mnras, 429,
  159

\bibitem[{{Grand} {et~al.}(2016){Grand}, {Springel}, {Kawata}, {Minchev},
  {S{\'a}nchez-Bl{\'a}zquez}, {G{\'o}mez}, {Marinacci}, {Pakmor}, \&
  {Campbell}}]{Grand2016}
{Grand}, R.~J.~J., {Springel}, V., {Kawata}, D., {et~al.} 2016, \mnras, 460,
  L94

\bibitem[{{Honma} {et~al.}(2012){Honma}, {Nagayama}, {Ando}, {Bushimata},
  {Choi}, {Handa}, {Hirota}, {Imai}, {Jike}, {Kim}, {Kameya}, {Kawaguchi},
  {Kobayashi}, {Kurayama}, {Kuji}, {Matsumoto}, {Manabe}, {Miyaji}, {Motogi},
  {Nakagawa}, {Nakanishi}, {Niinuma}, {Oh}, {Omodaka}, {Oyama}, {Sakai},
  {Sato}, {Sato}, {Shibata}, {Shiozaki}, {Sunada}, {Tamura}, {Ueno}, \&
  {Yamauchi}}]{Honma2012}
{Honma}, M., {Nagayama}, T., {Ando}, K., {et~al.} 2012, \pasj, 64, 136

\bibitem[{{Kuijken} \& {Tremaine}(1994)}]{Kuijken1994}
{Kuijken}, K. \& {Tremaine}, S. 1994, \apj, 421, 178

\bibitem[{{Lesh}(1968)}]{Lesh1968}
{Lesh}, J.~R. 1968, \apjs, 17, 371

\bibitem[{{Lindegren} {et~al.}(2016){Lindegren}, {Lammers}, {Bastian},
  {Hern{\'a}ndez}, {Klioner}, {Hobbs}, {Bombrun}, {Michalik}, {Ramos-Lerate},
  {Butkevich}, {Comoretto}, {Joliet}, {Holl}, {Hutton}, {Parsons},
  {Steidelm{\"u}ller}, {Abbas}, {Altmann}, {Andrei}, {Anton}, {Bach},
  {Barache}, {Becciani}, {Berthier}, {Bianchi}, {Biermann}, {Bouquillon},
  {Bourda}, {Br{\"u}semeister}, {Bucciarelli}, {Busonero}, {Carlucci},
  {Casta{\~n}eda}, {Charlot}, {Clotet}, {Crosta}, {Davidson}, {de Felice},
  {Drimmel}, {Fabricius}, {Fienga}, {Figueras}, {Fraile}, {Gai}, {Garralda},
  {Geyer}, {Gonz{\'a}lez-Vidal}, {Guerra}, {Hambly}, {Hauser}, {Jordan},
  {Lattanzi}, {Lenhardt}, {Liao}, {L{\"o}ffler}, {McMillan}, {Mignard}, {Mora},
  {Morbidelli}, {Portell}, {Riva}, {Sarasso}, {Serraller}, {Siddiqui}, {Smart},
  {Spagna}, {Stampa}, {Steele}, {Taris}, {Torra}, {van Reeven}, {Vecchiato},
  {Zschocke}, {de Bruijne}, {Gracia}, {Raison}, {Lister}, {Marchant},
  {Messineo}, {Soffel}, {Osorio}, {de Torres}, \& {O'Mullane}}]{Lindegren2016}
{Lindegren}, L., {Lammers}, U., {Bastian}, U., {et~al.} 2016, \aap, 595, A4

\bibitem[{{Michalik} {et~al.}(2015){Michalik}, {Lindegren}, \&
  {Hobbs}}]{Michalik2015}
{Michalik}, D., {Lindegren}, L., \& {Hobbs}, D. 2015, \aap, 574, A115

\bibitem[{{Monari} {et~al.}(2014){Monari}, {Helmi}, {Antoja}, \&
  {Steinmetz}}]{Monari2014}
{Monari}, G., {Helmi}, A., {Antoja}, T., \& {Steinmetz}, M. 2014, \aap, 569,
  A69

\bibitem[{{Pasetto} {et~al.}(2012){Pasetto}, {Grebel}, {Zwitter}, {Chiosi},
  {Bertelli}, {Bienayme}, {Seabroke}, {Bland-Hawthorn}, {Boeche}, {Gibson},
  {Gilmore}, {Munari}, {Navarro}, {Parker}, {Reid}, {Silviero}, \&
  {Steinmetz}}]{Pasetto2012}
{Pasetto}, S., {Grebel}, E.~K., {Zwitter}, T., {et~al.} 2012, \aap, 547, A70

\bibitem[{{Pickles} \& {Depagne}(2010)}]{Pickles2010}
{Pickles}, A. \& {Depagne}, {\'E}. 2010, \pasp, 122, 1437

\bibitem[{{Reid} {et~al.}(2014){Reid}, {Menten}, {Brunthaler}, {Zheng}, {Dame},
  {Xu}, {Wu}, {Zhang}, {Sanna}, {Sato}, {Hachisuka}, {Choi}, {Immer},
  {Moscadelli}, {Rygl}, \& {Bartkiewicz}}]{Reid2014}
{Reid}, M.~J., {Menten}, K.~M., {Brunthaler}, A., {et~al.} 2014, \apj, 783, 130

\bibitem[{{Rix} \& {Zaritsky}(1995)}]{Rix1995}
{Rix}, H.-W. \& {Zaritsky}, D. 1995, \apj, 447, 82

\bibitem[{{Rygl} {et~al.}(2012){Rygl}, {Brunthaler}, {Sanna}, {Menten}, {Reid},
  {van Langevelde}, {Honma}, {Torstensson}, \& {Fujisawa}}]{Rygl2012}
{Rygl}, K.~L.~J., {Brunthaler}, A., {Sanna}, A., {et~al.} 2012, \aap, 539, A79

\bibitem[{{Sch{\"o}nrich}(2012)}]{Schonrich2012}
{Sch{\"o}nrich}, R. 2012, \mnras, 427, 274

\bibitem[{{Siebert} {et~al.}(2011){Siebert}, {Famaey}, {Minchev}, {Seabroke},
  {Binney}, {Burnett}, {Freeman}, {Williams}, {Bienaym{\'e}}, {Bland-Hawthorn},
  {Campbell}, {Fulbright}, {Gibson}, {Gilmore}, {Grebel}, {Helmi}, {Munari},
  {Navarro}, {Parker}, {Reid}, {Siviero}, {Steinmetz}, {Watson}, {Wyse}, \&
  {Zwitter}}]{Siebert2011a}
{Siebert}, A., {Famaey}, B., {Minchev}, I., {et~al.} 2011, \mnras, 412, 2026

\bibitem[{{Widrow} {et~al.}(2014){Widrow}, {Barber}, {Chequers}, \&
  {Cheng}}]{Widrow2014}
{Widrow}, L.~M., {Barber}, J., {Chequers}, M.~H., \& {Cheng}, E. 2014, \mnras,
  440, 1971

\bibitem[{{Widrow} {et~al.}(2012){Widrow}, {Gardner}, {Yanny}, {Dodelson}, \&
  {Chen}}]{Widrow2012}
{Widrow}, L.~M., {Gardner}, S., {Yanny}, B., {Dodelson}, S., \& {Chen}, H.-Y.
  2012, \apjl, 750, L41

\end{thebibliography}

\end{document}